\pdfoutput=1 

\documentclass{pnastwo}
\usepackage{graphics,amssymb,amsmath,bm}

\newcommand{\ba}{\begin{eqnarray}}
\newcommand{\ea}{\end{eqnarray}}

\begin{document}

\title{A unified framework for  non-Brownian suspension flows and soft amorphous solids}
\author{Edan Lerner\affil{1}{Center for Soft Matter Research, Department of Physics,
New York University, New York, NY 10003, USA},
Gustavo D\"uring\affil{1}{Center for Soft Matter Research, Department of Physics,
New York University, New York, NY 10003, USA},
and Matthieu Wyart\affil{1}{Center for Soft Matter Research, Department of Physics,
New York University, New York, NY 10003, USA}.
}

\maketitle

\begin{article}

\begin{abstract}
While the rheology of non-Brownian suspensions in the dilute regime is well-understood,
their behavior in the dense limit remains mystifying. As the packing fraction
of particles increases, particle motion becomes more collective, leading to a growing
length scale 
and scaling properties in the rheology 
as the material approaches the jamming transition.
There is no accepted microscopic description of this phenomenon. However, in recent years it has
been understood that the elasticity of simple amorphous solids
 is governed by a critical point, the unjamming transition where the pressure vanishes,
 and where elastic properties display scaling and a diverging length scale.
 The correspondence between these two transitions is at present unclear.
 Here we show that for a simple model of dense flow, which we argue captures the essential physics near the jamming threshold,
 a formal analogy can be made between the rheology of the flow and the elasticity of simple networks. This 
 analogy leads to a new conceptual framework to relate microscopic structure
 to rheology.  It enables us to define and compute numerically normal modes and a density of 
 states. We find striking similarities between the density of states 
 in flow, and that of amorphous solids near unjamming:
 both display a plateau above some frequency scale $\omega^*\sim |z_{\rm c}-z|$, where $z$ 
 is the coordination of the network of particles in contact,  $z_{\rm c}=2D$ where $D$ is the spatial dimension.  
 However, a spectacular  difference appears: the density of states in flow displays a
 single mode at another frequency scale $\omega_{\rm min}<<\omega^*$  governing the divergence of the viscosity.
\end{abstract}

\keywords{suspensions | rheology}

% \abbreviations{KCM,~kinetically constrained model; FA,~Fredrickson-Andersen; 
% sFA,~`softened' FA;
% QPT,~quantum phase transition}

%\maketitle

\section{Introduction}

Suspensions are heterogeneous fluids containing solid particles. Their viscosity in the dilute regime
was computed early on by Einstein and Batchelor \cite{77Bat}. 
However, as the packing fraction $\phi$ increases, steric hindrance becomes dominant and particles 
move under stress in an increasingly coordinated way. For non-Brownian particles, the viscosity diverges as the suspension 
jams into an amorphous solid. This jamming transition %\cite{98LN} 
is a non-equilibrium
critical phenomena: the rheology displays scaling laws \cite{07OT, 10NVABZYGD, 09BJ, 95Dur,11BGP,10HBB,10TWRSH}
and a growing length scale \cite{07OT,10NVABZYGD,04Poul,11LAH}. Jamming occurs more generally in driven materials made 
of repulsive particles, such as in aerial granular flows \cite{99DD} with similar rheological phenomenology
\cite{09JPF}, and where large eddies are observed as jamming 
is approached \cite{04Poul}. This phenomenology bears similarity with that of the glass transition, 
where steric hindrance increases upon cooling, and where the dynamics becomes increasingly 
collective as relaxation times grow \cite{05BBBCMHLP}. In dense granular materials and in supercooled
liquids, the physical origin of collective dynamics and associated rheological phenomena 
remain elusive.

Recent progress has been made on a related problem, the {\it unjamming} transition 
where a solid made of repulsive soft particles is isotropically decompressed toward 
vanishing pressure \cite{03OSLN,05WNW,10Wya,05SLN,06ESHS,10MXL,10SQLN}. In this situation various properties of the 
amorphous solid, such as elasticity, transport and force propagation, display scaling with 
the distance from threshold \cite{03OSLN,09XVWLN,08WLKM} and are characterized by 
a diverging length scale \cite{05SLN,06ESHS}. The packing geometry also displays scaling: the 
average number of contacts or coordination $z$ of jammed particles, is observed to converge \cite{95Dur} to 
the minimal value allowing mechanical stability $z_{\rm c}$ derived by Maxwell \cite{64Max}. 
Theoretically, these observations can be unified by the realization that the vibrational spectrum 
of these simple amorphous solids must display a  cross-over at some threshold frequency \cite{05WNW}, distinct 
from Anderson localization, above which  vibrational modes are 
still extended but poorly transport energy \cite{10Wya}. This threshold frequency, corresponding to the 
so-called  boson peak ubiquitously observed in glasses \cite{81And}, % and believed 
%to play a key role  near the glass transition \cite{02GCGP}, 
is governed by coordination and
applied pressure \cite{05WSNW} and vanishes at the unjamming threshold. 

Under an imposed shear flow, materials near the jamming transition are strongly anisotropic and  particles receive a net force from  particles they are in contact with.  When an amorphous solid is decompressed toward its  unjamming transition however, configurations are isotropic and forces are balanced. Despite these significant differences, both transitions deal with the emergence or disappearance of rigidity, and
it is natural to seek a common description of these phenomena. Such a framework would unify 
the elastic and vibrational properties of amorphous solids, much studied in the context of granular materials, 
glasses \cite{81And} and gels of semi-flexible polymers \cite{05SPMLJ}, with the rheology 
of suspension  flows.  %observations, in particular the elastic properties of amorphous solidsranging from the anomalies  anomalous elasticity of solids that are about to unjam and the rheology of suspensions close to jamming. However,  Naive guesses on what this relationship may be strongly underestimates the rapid explosion of viscosity near jamming, as we shall see. Something is missing. 
Herein we build the foundation of this framework, by focusing on a model of a suspension flow  of 
non-Brownian hard particles where hydrodynamic interactions are neglected \cite{07OT}.  The relationship between this model and real suspensions is not obvious.  
Here we furnish numerical observations and physical arguments to support the contention that this model captures 
the relevant physics near jamming. The next step is to show that for this model, we can build a formal 
analogy between rheology and elasticity. This analogy allows us to simulate accurately and to 
measure with high precision key aspects of the particle organization, in particular the 
packing coordination. More fundamentally, it enables us to define and compute,
in flows of hard particles, normal modes and their density, which are the natural mathematical 
objects connecting geometry to rheology.  We find striking similarities between the jamming and 
the unjamming transitions:  the coordination converges to Maxwell's value with a non-trivial scaling
with pressure. Furthermore, the density of states displays a transition that is characteristic of
vibrational properties of amorphous solids.  However, a fundamental difference appears:  the density of states displays 
bi-scaling instead of the simple scaling found near unjamming, with one mode at 
low-frequency that strongly couples to shear, which is responsible for the rapid divergence in viscosity.  

\begin{figure}%\hspace{-0.7cm}
%\vspace{1.5cm}
\resizebox{7.7cm}{!}{\includegraphics{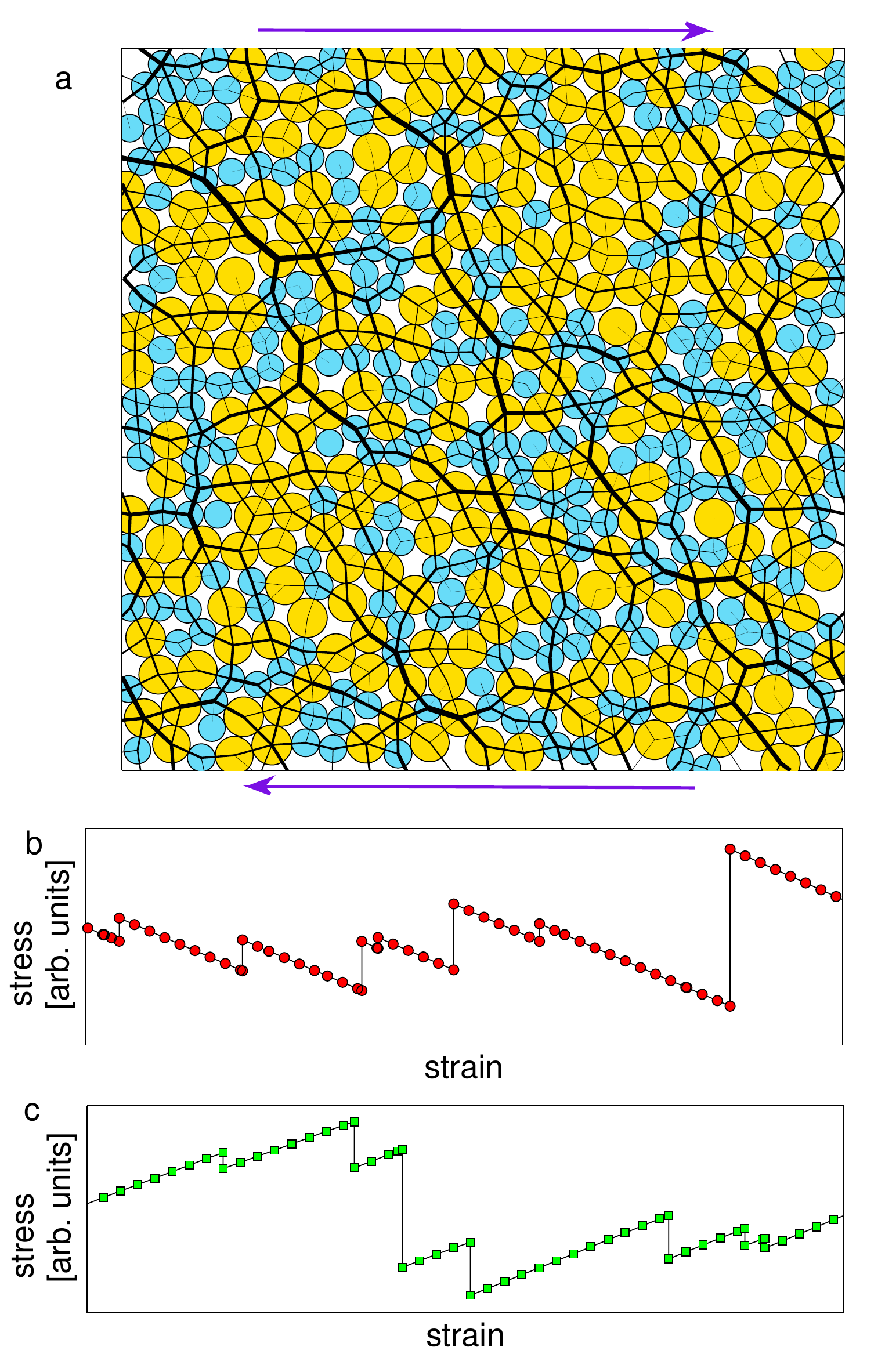}}
\caption{(a) Snapshot of a configuration visited under flow from simulations 
in two dimensions (see Methods Section for details). 
Small particles are in blue, large particles are in yellow.
The width of the black segments connecting particles is proportional to the contact force.
The arrows indicate the shear direction.
(b) Example of the evolution of the stress with 
time. Instantaneous jumps upward correspond to the creation of new 
contacts. Stress relaxes smoothly however within periods where contacts 
do not change. Interestingly, this situation is opposite to the plasticity 
of amorphous elastic solids, where stress loads continuously before relaxing 
by sudden plastic events, as illustrated in (c) for compressed 
soft elastic particles under quasi-static shear. \label{f1}}
\end{figure}

\section{Constitutive Relations}
Dimensional analysis implies that suspensions of over-damped hard particles are controlled by 
one dimensionless parameter \cite{11BGP},  the viscous number $I\equiv \dot\gamma\eta_0/\tilde p$
where $\tilde p$ is the particle pressure, $\eta_0$ is the fluid viscosity and $\dot{\gamma}$ is 
the strain rate. For convenience we shall consider its inverse, 
the normalized pressure $ p\equiv \tilde p/\dot\gamma\eta_0$. % that diverges near jamming.
%, since pressure as an obvious meaning both in elasticity and rheology. 
The steady-state rheology is then entirely determined by two constitutive relations that allow the computation
of the flow in various geometries \cite{11BGP,09JPF}:  $\phi(p)$ describes the dilatency 
of the material, and $\mu( p)\equiv \sigma/\tilde p $ its effective friction, where $\sigma$ is the shear stress.
Near jamming it is observed experimentally that:
\ba
\label{9}
\phi_{\rm c} - \phi(p) \propto 1/p^{\alpha},\\
\label{10}
\mu(p)-\mu_{\rm c}\propto 1/p^{\beta},
\ea
with $\alpha\approx1/2$ \cite{11BGP} % corresponding to a divergence of the viscosity $\eta\sim(\phi_c-\phi)^{-2}$,
and $\beta\approx 0.4$   \cite{11LAH,11BGP,09Pey}. % and a static friction $\mu_c=0.09$ in foams \cite{11LAH}. 
Currently these constitutive relations are phenomenological, and lack a microscopic theory. 

\begin{figure}
\vspace{-1.0cm}
\hspace{-0.50cm}
\resizebox{8.2cm}{!}{\includegraphics{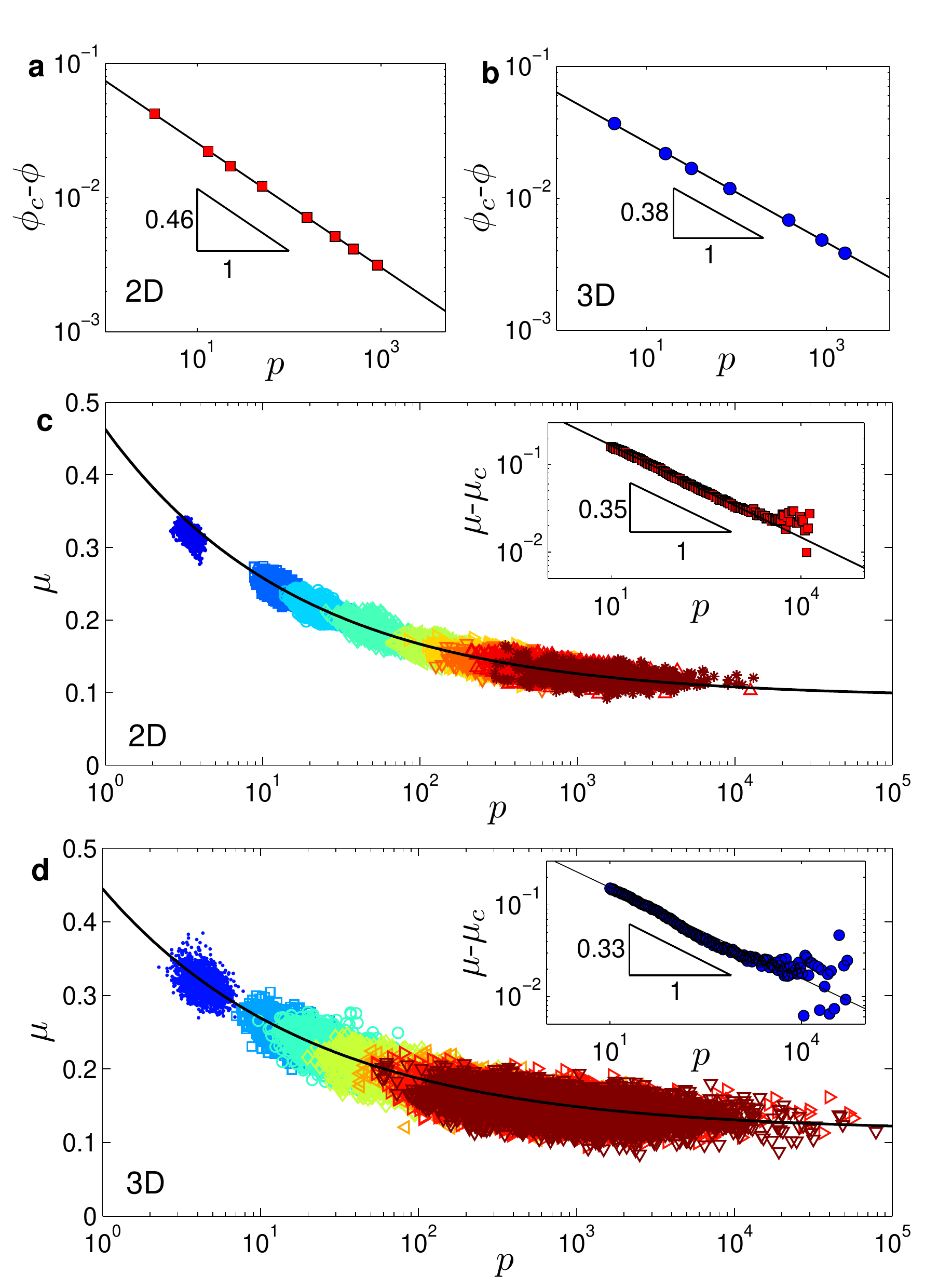}}
\caption{The  relation $\phi(p)$ within ASM is computed by averaging the normalized pressure $p$ at different packing fractions. This relation is well-captured 
by Eq.(\ref{9}) introduced to describe real suspension flows, with (a) $\phi_{\rm c}\approx0.842$ and $\alpha\approx0.46$ for $D=2$ 
and $N=4096$ particles, (b) $\phi_{\rm c}\approx0.647$ and $\alpha\approx0.38$ for $D=3$
and $N=1000$. The friction law $\mu(p)$ follows Eq.(\ref{10}), 
with (a) $\mu_{\rm c}\approx0.093$  and $\beta\approx0.35$ for $D=2$, and (b) $\mu_{\rm c}\approx0.115$ 
and $\beta\approx0.33$ for $D=3$.  The insets display $\mu-\mu_{\rm c}$ averaged 
at fixed pressure. The main plots show non-averaged data, and colors label 
the different packing fractions were measurements were made, growing from blue to red (see Methods Section). \label{f2}}
\end{figure}

\section{The Affine Solvent Model}
We consider a model of hard frictionless particles 
immersed in a viscous fluid, that we call the Affine Solvent Model (ASM). The dynamics is over-damped, and hydrodynamic 
interactions are neglected: the viscous drag on a particle is equal to the difference 
between the imposed velocity of the underlying fluid and the particle velocity, times a coefficient $\xi_0$.
The flow of the fluid phase is undisturbed by the particles and is chosen to be an affine shear of strain
rate $\dot\gamma$.   For hard particles at fixed packing fraction,  changing 
$\dot\gamma$ simply rescales time and the stress tensor must therefore be proportional to $\dot\gamma$ \cite{09LRC}.  
Note that for this model we can renormalize the pressure as $ p\equiv \tilde p/\dot\gamma\xi_0$ in two dimensions and $p\equiv \tilde p d/\dot\gamma\xi_0$in three dimensions,
where $d$ is taken to be the diameter of the small particles in our bi-disperse numerics.

% The rheological laws within this approximation of no hydrodynamical interactions were computed 
% numerically in two dimensions \cite{07OT, 11OT, 08Hat}. 
We constructed an event-driven simulation scheme (see \cite{11LDW} and the Methods 
Section below) based on an exact equation of motion derived below,
and on the actualization of the contact network, to simulate flow
while carefully extracting the geometry of configurations visited. 
In Fig.(\ref{f1}) we present a snapshot of a configuration from our simulations
in two-dimensions, and a typical stress-strain signal. A movie (Video S1) of our 
simulations can be found in the Supplementary Information.

Our results for the constitutive relations are shown in Fig.(\ref{f2}). 
We first note that fluctuations in pressure for a given $\phi$, proportional to the extent along the $p$-axis of the corresponding data-points cloud,  diverge
as jamming is approached, as expected from the observation that $\phi_c$ fluctuates in a finite size system \cite{11VVMOT}. This finite size effect creates a difficulty to 
accurately extract the exponent $\alpha$ that characterizes the dilatancy, see Eq.(\ref{9}).
Our simulation method does not allow for the study of larger systems,
which would improve our estimation of this exponent; nevertheless,
our results in two-dimensions are in good agreement with previous
work \cite{07OT, 11OT, 08Hat}. Our results for three dimensions
are similar to experimental observations: in three dimensions the friction law follows Eq.(\ref{10}) 
with $\beta\approx0.33$, % and $\mu_{c}=0.115$, 
whereas dilatency follows Eq.(\ref{9}) with  $\alpha\approx0.38$. These 
exponents are close to the experimentally observed exponents, supporting the claim that hydrodynamical 
interactions have weak effects, if any, on the critical behavior near jamming.
Note that the exponents are found to be similar in two and three dimensions,
raising the possibility that they are actually the same.

In sharp contrast with the fluctuations of pressure for a fixed packing fraction,
the fluctuations in various quantities considered \emph{as a function of pressure} are very limited,
and seem to remain regular throughout the entire range of pressures simulated, see Figs.(\ref{f2}),
(\ref{fig3}) and (\ref{f5}). In addition, we detect no finite size effects when computing
means as a function of pressure, see, for example, Fig.(\ref{f5}). Thus pressure 
is a much more suitable variable than packing fraction to quantify the distance from threshold 
and to study criticality in a finite system.

\section{Relationship between dynamics and packing geometry}
Within ASM the viscous drag force ${\vec F}_k$ acting on particle $k$ at 
position ${\vec R}_k$  is ${\vec F}_k=-\xi_0 ({\vec V_k}-{\vec V}^{\rm f}({\vec R}_k))$,
where ${\vec V_k}$ and ${\vec V}^{\rm f}({\vec R}_k)$ are the particle and fluid velocities respectively,
and $\xi_0$ characterizes the viscous drag.
This linear relation can be written for all particles in compact notation:
\begin{equation}
\label{1}
|F\rangle=-\xi_0(|V\rangle -|V^{\rm f}\rangle)\equiv -\xi_0 |V^{\rm na}\rangle,
\end{equation}
where the upper case ket notation $|X\rangle$ indicates a vector of dimension $ND$, $D$ being the
spatial dimension and $N$ the number of particles. 
$|V^{\rm na}\rangle\equiv |V\rangle -|V^{\rm f}\rangle$  is the so-called non-affine velocity. 
%This quantity is directly connected to the suspension viscosity $\eta$. Indeed the viscous power dissipated is $P\equiv\Omega \eta \dot\gamma^2\equiv-\langle F|V\rangle=\xi_0 ||V^{n.a}||^2$ where $\Omega$ is the volume of the system and Eq.(\ref{1}) was used to obtain the last equality. Thus $\eta\propto  ||V^{n.a}||^2/\dot\gamma^2$

In addition to the viscous drag, a force $f_{ik} {\vec n}_{ik}$ is exerted on the $k^{\rm th}$
particle from all particles $i$ which are in contact with particle $k$, where
${\vec n}_{ik}$ is the unit vector along ${\vec R}_k-{\vec R}_i$. 
In the viscous limit considered here, forces are balanced 
${\vec F}_k+\sum f_{ik}{\vec n}_{ik}=0$, where the sum is over all particles $i$ in 
contact with $k$. This equation can be written in compact notation:
\begin{equation}
\label{3}
|F\rangle+{\cal T} |f\rangle=0,
\end{equation}
where the lower case ket notation $|x\rangle$ indicates a vector of dimension $N_{\rm c}$,
the number of contacts. The linear operator ${\cal T}$  is of dimension $ ND\times N_{\rm c}$, 
its non-zero elements correspond to the vectors ${\vec n}_{ik}$, for 
particles $i$ and $k$ in contact. 

The last condition defining ASM is that particles cannot overlap. This condition is 
best expressed by considering the network of contacts, as illustrated in Fig.(\ref{f1}). 
This network rewires by instantaneous events where contacts open or close (see Video S1 in the S.I).
In between these discrete events, the network is conserved,
and the distance between particles in contact is fixed.
This condition implies that
\begin{equation}\label{moo}
({\vec V_k}-{\vec V_i})\cdot {\vec n}_{ik}=0\ ,
\end{equation}
for all contacts. The $N_{\rm c}$ linear constraints in the form of Eq.(\ref{moo})
(one constraint for each pair of particles in contact) can be written in compact notation as:
\begin{equation} \label{4}
{\cal S}|V\rangle = {\cal S}|V^{\rm f}\rangle + {\cal S}|V^{\rm na}\rangle=0  \quad \mbox{or} \quad 
{\cal S}|V^{\rm na}\rangle= -{\cal S}|V^{\rm f}\rangle,
\end{equation}
% or
% \begin{equation}
% {\cal S}|V^{\rm na}\rangle= -{\cal S}|V^{\rm f}\rangle,
% \end{equation}
where ${\cal S}$ is the $N_{\rm c}\times ND$ linear operator that computes 
the change in pairwise distances  $|\delta r\rangle$ between particles in contact,
induced by a displacement of the particles $|\delta R\rangle$: 
${\cal S}|\delta R\rangle=|\delta r\rangle$. A direct 
inspection of the elements of ${\cal S}$ indicates that it is the 
transpose of ${\cal T}$: ${\cal S}={\cal T}^t$. 

%%%%%%%%%%%%%%%%%%%%%  put this part in SI 
Now we may derive an expression for the viscosity. Operating with ${\cal S}$
on Eq.(\ref{1}), and using Eqs.(\ref{3}) and (\ref{4}), we obtain:
\begin{equation}
\label{5}
{\cal N} |f\rangle =-\xi_0 {\cal S}|V^{\rm f}\rangle
\end{equation}
where ${\cal N}={\cal S}{\cal T}={\cal S}{\cal S}^t$ is a $N_{\rm c}\times N_{\rm c}$ symmetric 
operator, which is generally invertible when the viscosity is finite (see below). ${\cal N}$ contains 
the information on the topology of the contact network (who is in contact with whom) 
and the contact orientations. ${\cal N}$ characterizes how a contact force field $|f\rangle$ is 
unbalanced $\langle  f|{\cal N}|f\rangle=\langle  f|{\cal T}^t{\cal T}|f\rangle = \langle F|F\rangle$. The term ${\cal S}|V^{\rm f}\rangle$ on the right 
hand-side of Eq.(\ref{5}) is not singular near jamming, as it is the 
rate at which the contacts lengths would change (generating gaps or overlaps 
between particles) if particles were following the affine flow of the fluid.
One can thus introduce the notation 
${\cal S}|V^{\rm f}\rangle\equiv \dot\gamma |\gamma\rangle$. For a pure shear in the $(x,y)$ plane, the components of 
$|\gamma\rangle$ are $\gamma_{ij}=r_{ij} ({\vec n}_{ij}\cdot  {\vec e}_{x})({\vec n}_{ij}\cdot  {\vec e}_{y})$, where $r_{ij}$ is the distance between particles $i$ and $j$. Inverting Eq.(\ref{5}) and using this notation we derive 
an expression for the contact forces:
\begin{equation}
\label{6}
|f\rangle =-\xi_0 \dot\gamma {\cal N}^{-1} |\gamma\rangle
\end{equation}
This result, together with Eqs.(\ref{1}) and (\ref{3}) yields an expression for the dynamics:
$ |V^{\rm na}\rangle=-\dot\gamma{\cal S}^t{\cal N}^{-1}|\gamma\rangle$. 
The shear stress $\sigma$ carried by the particles is related to the contact 
forces by the relation \cite{91AT} $\sigma\equiv-{\vec e}_{x}\cdot (\sum_{ij}f_{ij} r_{ij}{\vec n}_{ij}\otimes {\vec  n}_{ij})\cdot{\vec e}_{y}/\Omega=-\langle f|\gamma\rangle/\Omega$, 
where $\Omega$ is the volume of the system.
%%%%%%%%%
We thus obtain from Eq.(\ref{6}) the suspension viscosity $\eta$:
\begin{equation}
\label{7}
\eta = \frac{\xi_0}{\Omega} \langle \gamma|{\cal N}^{-1} |\gamma\rangle
\end{equation}
Eq.(\ref{7}) is a key result, as it shows that all the singularities that appear 
near jamming within ASM are contained in a single operator, ${\cal N}$, which 
connects rheology to geometry. 

\section{Analogy between elasticity and rheology}
The operator ${\cal N}$ bears similarities with the well-studied
stiffness matrix ${\cal M}$ that plays a central role in elasticity.
By definition, the quadratic expansion of the energy in any elastic 
system is $\delta E=\langle \delta R| {\cal M}|\delta R\rangle/2$.
Consider the case of an elastic network of un-stretched springs of unit stiffness.
Labeling a spring by $\alpha$, the energy can be expressed in terms of the spring elongation 
$\delta r_\alpha$, namely:
\begin{equation}
\label{8}
\delta E=\sum_\alpha \frac{1}{2} \delta{r_\alpha}^2\equiv
\frac{1}{2}\langle\delta r|\delta r\rangle=\frac{1}{2}\langle{\cal S}
\delta R|{\cal S}\delta R\rangle=\frac{1}{2}\langle\delta R|{\cal S}^t{\cal S}|\delta R\rangle
\end{equation}
Together with the definition of ${\cal M}$, Eq.(\ref{8}) implies 
that ${\cal M}={\cal S}^t{\cal S}$. This expression is closely related 
to ${\cal N}={\cal S}{\cal S}^t$, building a connection between 
rheological properties of suspended hard particles 
and the elasticity of spring networks with identical geometry.
In particular, the spectra of ${\cal M}$
and ${\cal N}$ are identical for positive eigenvalues 
$\lambda\equiv \omega^2$, where $\omega$ is referred to as the mode 
frequency. To see this let us assume that $|\delta r(\omega)\rangle$ is a normalized eigenvector of ${\cal N}$, then
\begin{equation}
{\cal S}{\cal S}^t |\delta r(\omega)\rangle=\omega^2 |\delta r(\omega)\rangle\ .
\end{equation}
Operating on the above equation with ${\cal S}^t$ we find:
\begin{equation}
{\cal S}^t{\cal S}{\cal S}^t |\delta r(\omega)\rangle\equiv {\cal M}{\cal S}^t |\delta r(\omega)\rangle=\omega^2 {\cal S}^t |\delta r(\omega)\rangle\ ,
\end{equation}
implying that ${\cal S}^t |\delta r(\omega)\rangle$ is an eigenvector of ${\cal M}$ with the same eigenvalue $\omega^2$. 
Thus, there is a one-to-one correspondence between eigenvectors of 
${\cal M}$ and ${\cal N}$ for positive eigenvalues.

%, and that the  eigenvectors $|\delta r_\omega\rangle$ and $|\delta R_\omega\rangle$ of ${\cal N}$ and ${\cal M}$ respectively satisfy  $\omega |\delta r_\omega\rangle= {\cal S}|\delta R_\omega\rangle$. 
This observation suggests that the analytical techniques that were
developed to study elasticity in random media apply as well to rheology 
of suspensions, at least within ASM. Here we shall take another route by studying
empirically the spectrum of ${\cal N}$, identifying what the spectral signatures 
of the singularities are in the rheological laws, 
and compare those with the known singularity 
in the spectrum of ${\cal M}$ near the unjamming transition.  

\begin{figure}
\vspace{-0.6cm}
\resizebox{7.2cm}{!}{\includegraphics{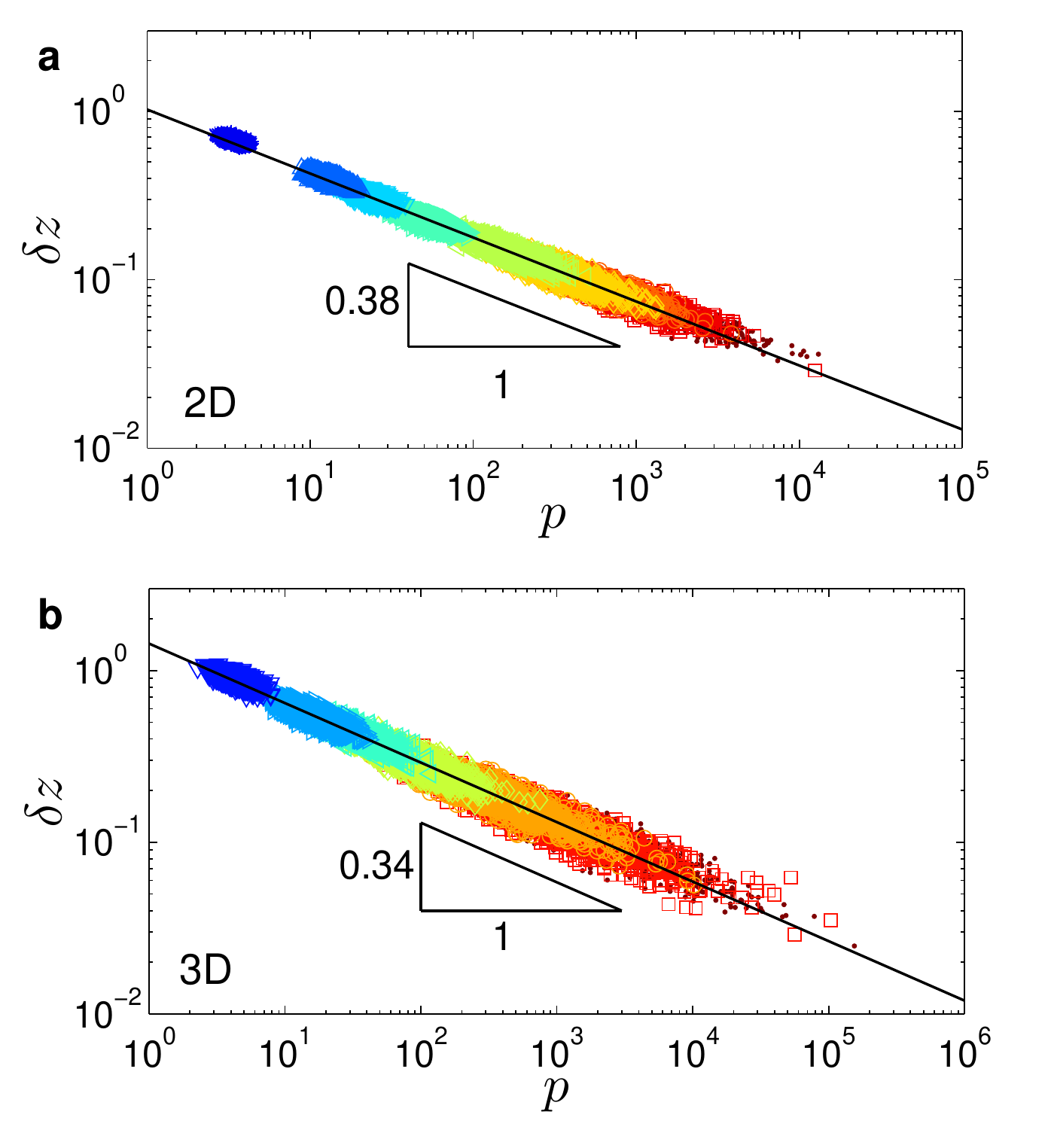}}
\caption{The coordination displays scaling near jamming $\delta z\equiv z_{\rm c}-z\sim 1/p^{\delta}$ 
in (a) two and (b) three dimensions. Particles barely connected to the contact network
(making less than two contacts with their surrounding), the so-called 
rattlers, are removed from the analysis.  Colors label different packing fractions (see Methods Section), growing from blue to red,
and each data point pertains to a single configuration. 
% Note that it is apparent from Fig.(\ref{fig3}) that finite size 
% fluctuations of $p$ at fixed $\phi$ diverge as jamming is approached.
% However, fluctuations of $z$ at fixed $p$ remain 
% very limited. Thus pressure is a much  better variable than $\phi$ to quantify the distance from threshold 
% and to study criticality in a finite system.
\label{fig3}}
\end{figure}

\begin{figure*}
\begin{center}\label{f4}
%\resizebox{7in}{!}{\includegraphics{fig4_4.pdf}}
\resizebox{6.8in}{!}{\includegraphics{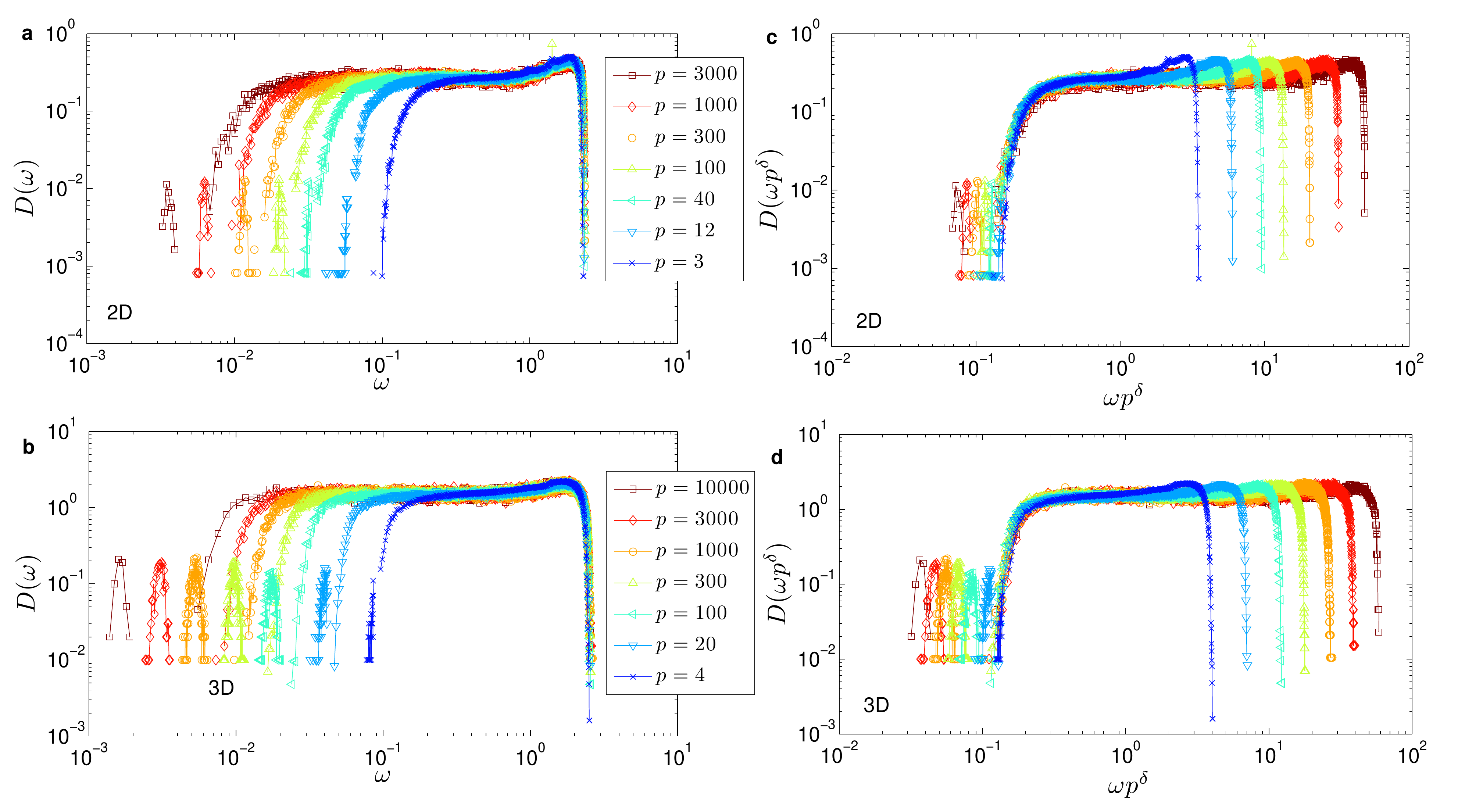}}
\caption{Spectral analysis of the operator ${\cal N}$ governing flow as 
jamming is approached. The density of states $D(\omega)$  averaged over 500 configurations for
each pressure indicated in the legend, in (a) two and (b) three dimensions. The amplitude of the peak at low-frequency was rescaled to make it visible.
Since it consists of one mode only, this amplitude vanishes in the thermodynamic limit. 
If the frequency axis is rescaled by the excess coordination $p^\delta\sim\delta z$, 
as in (c,d), the emergence of a plateau in $D(\omega)$ collapses, indicating that the frequency at which this plateau appears follows $\omega^*\sim \delta z$.}
\end{center}
\end{figure*}

\section{Microscopic analysis of flow configurations}
A key variable that characterizes steric hindrance is the average 
number of contacts per particle, or the coordination $z=2N_{\rm c}/N$. This intuition is 
substantiated by Eq.(\ref{7}), which  indicates that the viscosity diverges 
when det$({\cal N})=0$. Since ${\cal N}={\cal S}{\cal S}^t$, 
${\cal N}$ must display zero-modes if ${\cal S}^t$ does, which must 
occur if $N_{\rm c}>ND$, as ${\cal S}^t$ is of dimension $ND\times N_{\rm c}$.
For the coordination this argument implies that jamming must occur if
$z>2D\equiv z_{\rm c}$, a necessary condition for rigidity derived by 
Maxwell \cite{64Max}. Earlier numerical work \cite{09HB}
considered the scaling of the coordination with $\phi$,
where it was found that for $\phi < \phi_c$, $z_0 - z\sim \phi_c - \phi$ with $z_0 = 3.8$.
However, this analysis is hampered
by the rather large fluctuations of $\phi_c$ in  finite size 
systems \cite{11VVMOT}, in addition to errors originating from the inclusion of rattlers. 
Our numerical procedure allows for a 
careful measurement of the coordination, displayed in Fig.(\ref{fig3}). We find instead:
\begin{equation}
\label{11}
 \delta z\equiv z_c-z\sim 1/p^\delta\ ,
\end{equation}
with $z_c = 2D$, $\delta \approx 0.38$ and $\delta\approx0.34$ in two and three dimensions respectively. 

\section{Density of states}
The symmetric operator ${\cal N}$ is readily obtained from the contact network,
and its spectrum is computed numerically. The frequency spectra, or 
densities of states $D(\omega)$ defined as the number of modes per unit 
frequency per particle, are shown in Fig.(\ref{f4}) as jamming is approached.
They consist of two structures: a plateau of modes appearing above some 
frequency threshold $\omega^*$, and a gap at lower frequency, containing 
only one isolated mode of frequency $\omega_{\rm min}$. Strikingly, this plateau 
is also present in the vibrational spectrum of simple amorphous 
solids \cite{03OSLN,05SLN}, where its onset frequency scales as $z-z_c$ \cite{05WNW,10Wya}. 
To determine if this scaling law holds in configurations visited in flows as well,
we rescale the frequency axis by $\delta z\sim 1/p^\delta$.  
Fig.(\ref{f4}) displays a striking collapse 
of the plateau in the density of states, emphasizing the strong connection between 
elasticity near unjamming and the rheology of dense flows. 

\begin{figure*}
\begin{center}\label{f5}
\resizebox{5.5in}{!}{\includegraphics{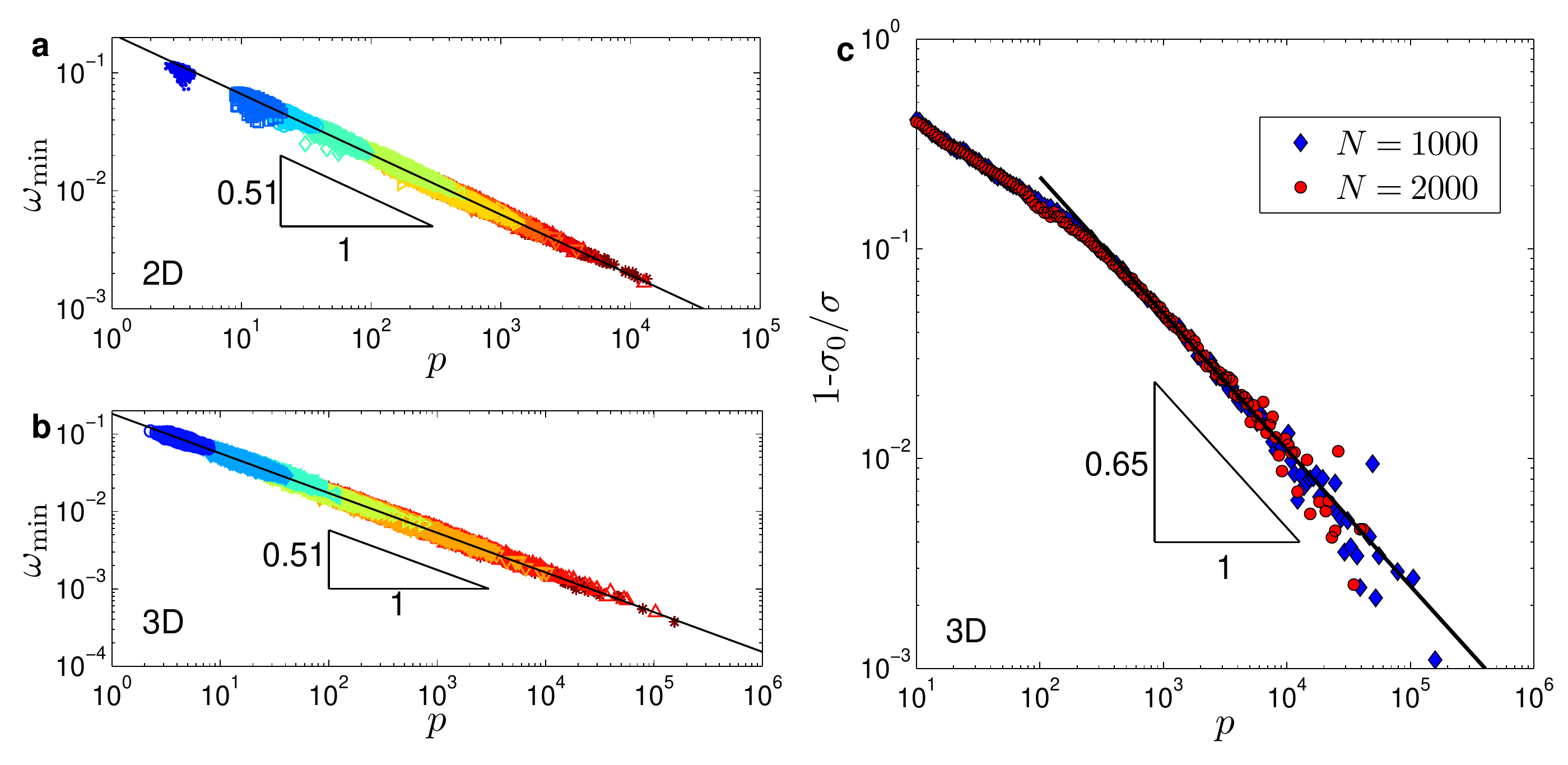}}
\vspace{-0.3cm}
\caption{Scaling of the lowest frequency $\omega_{\rm min}$ vs.~normalized
pressure $p$ in (a) 2 and (b) 3 dimensions.
This modes dominates the rheology near jamming, as can be seen 
by plotting the relative contribution $(\sigma-\sigma_0)/\sigma$ of all the other 
modes to the stress, that vanishes as $p^{-0.65}$. The relative contribution is found to 
be independent of $N$.}
\end{center}
\end{figure*}

The sole exception is the isolated mode at low-frequency that does not collapse with 
this rescaling. Rather, its frequency scales as $\omega_{\rm min}\sim 1/p^{\epsilon}$ 
with $\epsilon\approx0.51$ independently of the spatial dimension, 
as shown in Fig.(\ref{f5}). Such bi-scaling is a new feature of flow 
not present in the elasticity of solids near unjamming, and it is of critical importance 
for the rheology. 
% Theoretical attempts to describe flow by assuming a perfect analogy 
% with the unjamming transition are therefore incorrect \cite{11Tig}. 

 Using Eq.(\ref{7}) we can separate the contributions of the first mode $\sigma_0$ and of the plateau $\sigma-\sigma_0$ 
to the shear stress $\sigma$. Denoting the lowest-frequency mode as $|\delta r_0\rangle$,
and the rest of the modes by $|\delta r_\omega\rangle$, we obtain:
\ba
\label{13}
\sigma =  \frac{\xi_0\dot{\gamma}}{\Omega}\frac{\langle \gamma|\delta r_0\rangle^2}{\omega_{\rm min}^2 }
+\frac{\xi_0\dot{\gamma}}{\Omega}\sum_{\omega>\omega^*} \frac{\langle\gamma|\delta r_\omega\rangle^2}{\omega^2}
\ea
We find that the first term of Eq.(\ref{13}), denoted $\sigma_0$, dominates stress
near jamming: $\sigma \rightarrow \sigma_0$.
This stems from the fact that the first eigenvector has a very strong projection on a shear,
that does not vanish as the number of particle 
increases: $\langle \gamma|\delta r_0\rangle/(||\gamma||\!\!\times\!\! ||  \delta r_0||)\!\approx\! 0.15$ in
three dimensions. Using that $||\gamma||\!\sim\! N$ and that $||\delta r_0||\!=\!1$  leads 
to $\sigma_0\!\sim\! 1/\omega_{\rm min}^2$. Since $\sigma_0\!\sim\!\sigma\!\sim\! p$ we obtain
$\omega_{\rm min}\!\sim\! 1/\sqrt p$ or $\epsilon=1/2$, very close to the observation $\epsilon\!\approx\!0.51$.

Note that the strong coupling between $|\delta r_0\rangle$ and $|\gamma\rangle$ implies, together with Eq.(\ref{6}), that $|f\rangle/||f||\rightarrow |\delta r_0\rangle$ as $p\rightarrow \infty$.
Thus Fig.(1a) and the movie in S.I., that display respectively an example of $|f\rangle$ and its evolution under shear very close to threshold, are very accurate representation of $|\delta r_0\rangle$. 

The relative contribution of the rest of the density of states, corresponding to the second term in Eq.(\ref{13}),
vanishes as $(\sigma-\sigma_0)/\sigma \sim p^{-0.65}$, as 
shown in Fig.(\ref{f5}). This exponent can also be rationalized by 
a simple scaling estimate, similar to arguments previously introduced for isotropic configurations
near the unjamming transition \cite{08WLKM,11Tig}. Assuming that the modes $|\delta r_\omega\rangle$ 
forming the plateau in $D(\omega)$ have random orientations with respect to a shear
\cite{08WLKM} leads to $\langle\gamma|\delta r_\omega\rangle^2\sim O(1)$. Thus
\begin{eqnarray}
\frac{\sigma-\sigma_0}{\xi_0\dot{\gamma}} & \equiv &\frac{1}{\Omega} \sum_{\omega>\omega^*}
\frac{\langle\gamma|\delta r_\omega\rangle^2}{\omega^2}\sim 
\frac{1}{\Omega}\sum_{\omega>\omega^*} \frac{\langle\delta r(\omega)|\gamma\rangle^2}{\omega^2}  \nonumber \\
& \sim & \int\limits_{\omega>\omega^*}\frac{D(\omega)d\omega}{\omega^2}
 \sim \int\limits_{\omega>\omega^*}\frac{d\omega}{\omega^2}
\sim 1/\omega^*\sim 1/\delta z \ , \nonumber 
\end{eqnarray}
implying $(\sigma-\sigma_0)/\sigma \sim \delta z/\sigma\sim
\delta z/p\sim p^{\delta -1}$ which is indeed $p^{-0.65}$ in three 
dimensions. Thus most of the spectrum contributes less and less to the 
divergence of stress as jamming is approached.
This contribution may nevertheless play an important role in the corrections to scaling, such 
as those leading to a varying friction law in Eq.(\ref{10}). We shall explore this
possibility in future work. 

\section{Discussion}

We have shown that the rheological properties of ASM flows are described 
by a single operator ${\cal N}$, which is closely connected to the stiffness matrix of elastic networks.
This result allows for a novel characterization of flow in terms of the spectrum of a single operator ${\cal N}$. 
This spectrum presents remarkable features: it displays the plateau of modes controlling the anomalous elastic properties of amorphous solids
near the unjamming transition, but also one mode at low-frequency responsible for the sharp divergence of the viscosity.
Future work necessary to build a description of flow near jamming should explain (i) what configurations can generate such a bi-scaling spectrum  (ii) why such configurations are selected by the dynamic and (iii) what fixes the relationship between $\phi$ and $z$. 

It is likely that such a description will be of mean-field character. Near the unjamming transition the fact that a frequency scale $\omega^*$ scales linearly with the excess coordination $\delta z$ in any dimension reflects the fact that spatial fluctuations of coordination 
are weak, and irrelevant. The unjamming transition is in some sense a mean-field version of rigidity percolation
where bonds are deposited randomly on a lattice and where fluctuations are important \cite{85GT},
as a mean-field description of the latter gives the correct elastic behavior near unjamming \cite{10Wya}.
Our finding that some frequency scale $\omega^*$ also scales like $\delta z$ in ASM flows %thus 
suggests that spatial fluctuations of coordination are irrelevant in dense suspensions too. This is also supported by our observation that 
more generally, exponents appear to depend weakly, if at all, on spatial dimension.  

To conclude, we discuss the generality of our results.  Constitutive relations in ASM appear quantitatively similar to experiments in real suspensions.
This result might seem surprising at first glance, since the fluid is certainly strongly disturbed
by flow near jamming, unlike what ASM assumes. We speculate why these two problems may fall in the same universality class.
In real flow dissipation is dominated by the viscous friction resulting from lubrication  between neighboring particles. The  corresponding tangential friction coefficient has a weak (logarithmic) dependence on the gap
between particles, whereas the normal friction coefficient diverges as the inverse of the gap.  However evidences \cite{95BM,09Pey} support that due to the finite elasticity and/or  roughness
of the particles, real contacts are eventually formed \cite{95BM,09Pey}. If so, one expects that the divergence of the normal friction coefficient will have a cut-off, and
that the viscosity will be proportional to the square of the relative velocities between particles. In ASM,
the viscosity is proportional to the square of the non-affine velocities. However, we have checked that both
quantities are approximatively proportional to each other as jamming is approached, supporting that this model
is appropriate to describe the scaling properties of suspension flows near jamming. Further support comes
from a recent work on the viscoelasticity
of amorphous solids near unjamming \cite{11Tig}, showing that variations in the dissipation mechanism need not alter the scaling relations for the viscoelastic properties.

\begin{acknowledgments}
We  thank Y. Elmatad, A. Grosberg, P. Hohenberg, D. Pine, E. Vanden-Eijnden and anonymous referees for constructive comments on the manuscript. This work has been supported by
the Sloan Fellowship, NSF DMR-1105387, and Petroleum Research Fund \#52031-DNI9.
\end{acknowledgments}

\vspace{-0.3cm}
\section{Methods}
The simulations are based on the ASM equations of motion, which prescribes a velocity to each particle for a given configuration.
We propagate the system forward in time
over small time steps, while carefully monitoring the formation of new 
contacts, or the destruction of existing contacts. For a complete description of the simulation
employed see \cite{11LDW}. 

We simulated systems of $N_{\rm 2D} = 4096$ particles in two dimensions, at packing fractions 
$\phi_{\rm 2D} = 0.82$, 0.825, 0.83, 0.835, 0.837, 0.838, 0.839, and 0.840. 
We also simulated systems of $N_{\rm 3D} = 1000$ and 2000 particles in three dimensions, at packing
fractions $\phi_{\rm 3D} = 0.61$, 0.625, 0.63, 0.635, 0.640, 0.642, and 0.643. In all simulations we used
a square box with Lees-Edwards periodic boundary conditions \cite{91AT} that allows for homogeneous shear flow profiles
throughout the simulation cell. Our systems consist of a 50:50 mixture of small and large
particles, where the ratio of the diameters $d$ of the 
large and small particles is $d_{\rm large}/d_{\rm small} = 1.4$. We add a slight poly-dispersity of 
3\% in the particle sizes to avoid hexagonal patches in the two-dimensional systems \cite{11LDW}. 
Normal modes are calculated with the LAPACK software package (http://www.netlib.org/lapack/).

\bibliographystyle{ieeetr}

\end{article}

\end{document}